\documentclass[11pt]{article}
\usepackage[margin=2.5cm]{geometry}
\usepackage{amsmath,amssymb,graphicx,booktabs}
\usepackage[hidelinks]{hyperref}
\usepackage{microtype}
\usepackage{authblk}
\usepackage{caption}
\title{Does a Higher Frame Rate Improve the Measurement of \texorpdfstring{$g$}{g}? Precision and Accuracy in Video Analysis}
\author[1,2]{Mauricio Echiburu Fuenzalida\thanks{mauricio.echiburu@userena.cl}}
\author[3]{Nicolás Fernández-Astudillo\thanks{nicolas.fernandez@upla.cl}}
\affil[1]{Department of Physics, Faculty of Sciences, Universidad de La Serena}
\affil[2]{Bachelor's Programme in Astronomy, Faculty of Engineering and Architecture, Universidad Central}
\affil[3]{Physics Education Laboratory (DFIS-UPLA), Faculty of Natural and Exact Sciences, Universidad de Playa Ancha}
\date{September 11, 2026}
\begin{document}
\maketitle

\begin{abstract}
This study investigates the influence of frame rate, expressed in frames per second (FPS), on the experimental estimation of the acceleration due to gravity, $g$, using video analysis with Tracker. Two different dynamical systems were analysed: free fall and a simple pendulum. Starting from original recordings acquired at 120 FPS, equivalent 60, 30, and 10 FPS series were generated by uniform temporal decimation, preserving the same physical trajectory within each realisation. The analysis included ten independent realisations for free fall and eight for the pendulum. For each condition, the mean estimate of $g$, the standard deviation across realisations, the relative error with respect to $g_{\mathrm{ref}}=9.81\,\mathrm{m\,s^{-2}}$, and the RMSE were calculated.

For free fall, the 120 FPS condition exhibited the lowest dispersion and RMSE, whereas the mean estimates obtained at 10 and 30 FPS were closer to the reference value. For the simple pendulum, by contrast, the mean value of $g$, dispersion, relative error, and RMSE remained essentially unchanged between 10 and 120 FPS. An additional analysis of the 12 possible subsampling phases at 10 FPS showed that this stability does not critically depend on a particular choice of decimation phase and that the variability associated with phase is small compared with the dispersion observed across realisations.

The results show that a higher temporal sampling rate does not necessarily provide both greater precision and closer agreement with the reference value. The influence of FPS depends on the temporal characteristics of the system and on the procedure used to estimate the physical parameter. In particular, a higher temporal sampling density may reduce dispersion or increase robustness to sampling without necessarily yielding an estimate closer to the expected value.
\end{abstract}


\section{Introduction}
Video analysis has become a widely used tool for studying mechanical motion from experimental recordings. Its main advantage is that it allows spatial and temporal information to be obtained directly from an image sequence and compared with physical models. Among the available tools, Tracker has been used in numerous mechanics experiments, including free fall, oscillatory motion, and pendulum systems (Bhakat et al., 2024; Brown \& Cox, 2009; Grimaldi et al., 2026).

The measurement of the acceleration due to gravity, $g$, provides a particularly suitable case for evaluating the performance of such procedures. The same physical quantity can be determined using different dynamical systems and therefore from different characteristics of motion. In free fall, $g$ is obtained from the curvature of the position-time trajectory, whereas in a simple pendulum it is determined primarily from the measurement of the period. Video analysis applied to both procedures has yielded results compatible with expected values of $g$, although with different levels of dispersion and experimental error (Grimaldi et al., 2026; Pacala \& Mendaño, 2025; Pacala \& Pili, 2023).

A fundamental characteristic of any video recording is its temporal sampling rate, commonly expressed in frames per second (FPS). For a recording of fixed duration, increasing the FPS increases the number of available observations and allows the motion to be described with a higher temporal sampling density. This may be particularly relevant when the physical parameter of interest is obtained by fitting the temporal evolution of the system, since a larger number of points may reduce the sensitivity of the estimate to localisation errors or to the particular selection of observations used (Ramli et al., 2016).

However, having a larger number of observations does not necessarily imply that the estimated value of a physical quantity will be closer to its reference value. To analyse this issue properly, precision and accuracy must be distinguished. In this work, precision is understood as the dispersion observed among measurements obtained from independent realisations, whereas accuracy is discussed in terms of how close the results are to the reference value adopted for $g$. According to this distinction, a set of measurements may exhibit low dispersion while at the same time being systematically shifted from the expected value (Joint Committee for Guides in Metrology, 2012).

Although numerous studies have used Tracker to analyse mechanical motion and determine physical parameters (Bhakat et al., 2024; Brown \& Cox, 2009; Brown et al., 2026; Grimaldi et al., 2026; Pacala \& Mendaño, 2025), much of this work has focused on demonstrating the feasibility of video analysis, testing physical models, or comparing experimental procedures. Frame rate is usually considered mainly from the standpoint of the temporal resolution of the recording. By contrast, the relationship between frame rate and the simultaneous evolution of precision, closeness to the reference value, and the overall error of the estimated parameter has received comparatively less attention.

The influence of FPS also need not be independent of the temporal characteristics of the phenomenon under study. In a short-duration transient motion such as free fall, reducing the temporal sampling rate directly decreases the number of observations used to characterise a non-repeating trajectory. In periodic motion, by contrast, the temporal information required to determine the period appears repeatedly over several cycles. Moreover, in the latter case, a substantial reduction in FPS may introduce an additional dependence on the subsampling phase, since different starting points of the same decimation can select different regions of the oscillation cycle.

In this work, the influence of frame rate on the measurement of $g$ is experimentally studied using two dynamical systems: free fall and a simple pendulum. From original recordings obtained at 120 FPS, equivalent series at 60, 30, and 10 FPS are generated by uniform temporal decimation. This procedure changes the temporal sampling density while preserving, within each realisation, the same physical trajectory and prevents differences between FPS conditions from being attributed to changes in the initial conditions or experimental setup.

To characterise the effect of reducing FPS, the mean estimate of $g$, the standard deviation across realisations, the relative error with respect to a reference value, and the RMSE are considered jointly. This makes it possible to distinguish changes in precision, changes in the closeness of the mean to the expected value, and changes in the overall performance of the measurements. For the pendulum, a sensitivity analysis of the subsampling phase at 10 FPS is also included to determine whether the stability of the estimate depends on a particular selection of temporal points.

The central objective of this study is to determine how the precision, accuracy, and overall error of the measurement of $g$ vary when the frame rate is changed in two different experimental procedures. In particular, the hypothesis that increasing FPS simultaneously leads to more precise measurements and measurements closer to the reference value is tested. The comparison between free fall and the simple pendulum also makes it possible to evaluate whether this relationship depends on the transient or periodic nature of the system and on the procedure used to estimate the physical parameter.

\section{Methodology}
\subsection{General design}
\begin{figure}[ht]
\centering
\includegraphics[width=0.58\linewidth]{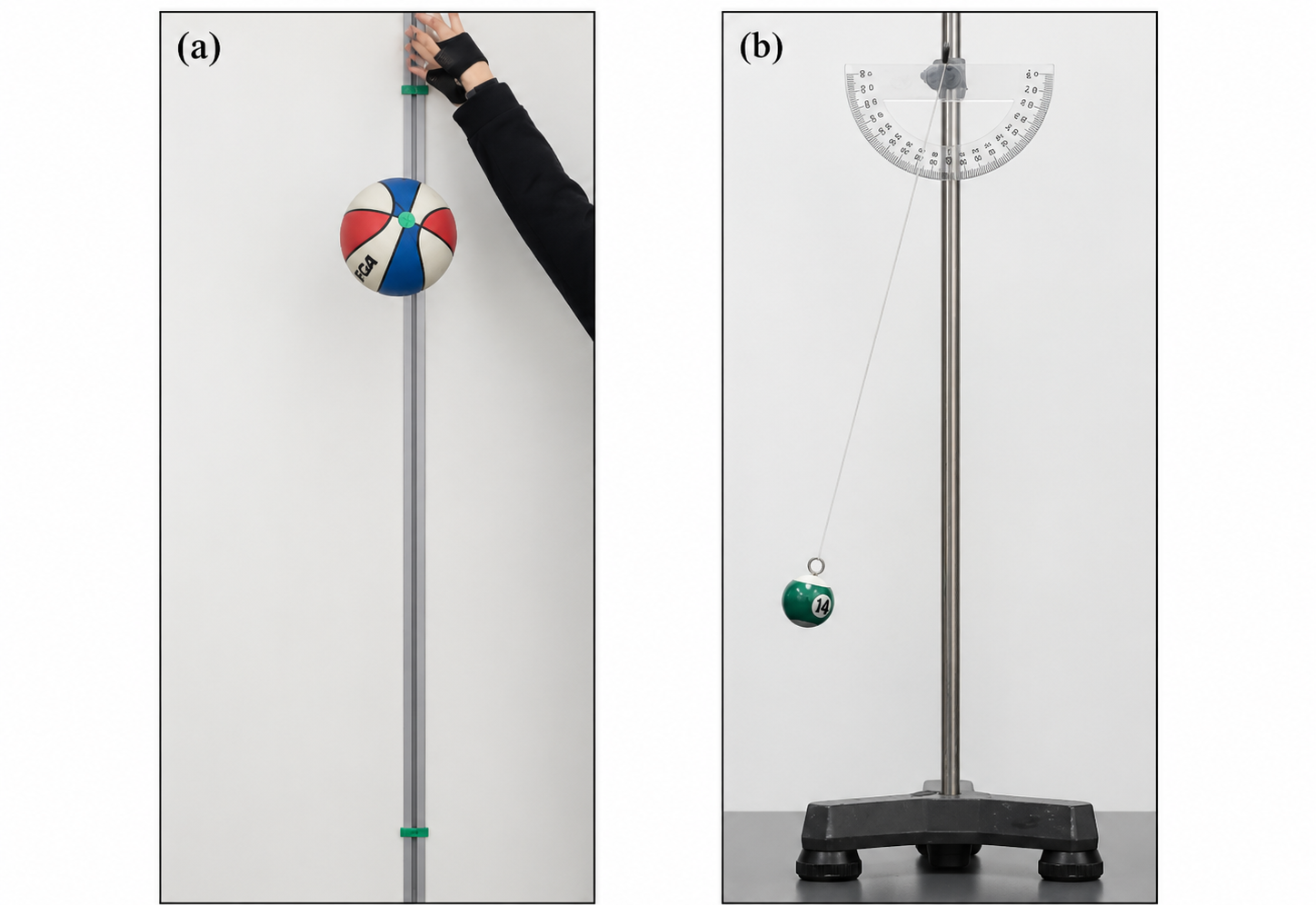}
\caption{Experimental setups used to measure $g$: (a) free fall of a ball travelling a vertical distance of 2.0 m between two reference marks and (b) simple pendulum with a 0.50 m string.}
\label{fig:setups}
\end{figure}

The study was designed to evaluate how frame rate affects the experimental measurement of the acceleration due to gravity, $g$, using two different dynamical systems: free fall and a simple pendulum. Figure~\ref{fig:setups} shows the two experimental setups used in the study. In the free-fall experiment, the vertical motion of a ball was recorded along a fixed spatial reference, whereas the pendulum experiment used a mass suspended from a string of known length and a support with an angular reference.

In both experiments, a 120 FPS recording was used as the original condition. Equivalent versions at 60, 30, and 10 FPS were constructed from these data by controlled reduction of the temporal sampling rate.

The purpose of this procedure was to compare different temporal sampling densities while preserving, within each realisation, the same original physical trajectory. The main analysis used a single decimation phase for each temporal sampling rate, defined from the first point of each record. In addition, a sensitivity analysis was performed for the pendulum to determine whether the results obtained at 10 FPS depended on the phase used for temporal subsampling.

\subsection{Reduction of the temporal sampling rate}
Each experimental realisation was originally recorded at 120 FPS. The lower-frequency conditions do not correspond to new recordings; instead, they were obtained from each original record through a uniform temporal decimation procedure.

For the main analysis, the equivalent 60, 30, and 10 FPS series were constructed by always retaining the first point of the record and subsequently selecting one out of every 2, 4, and 12 points, respectively. If the original series is represented as
\begin{equation}
\{(t_i,x_i)\}_{i=1}^{N},
\end{equation}
the reduced series were obtained by selecting the indices corresponding to the decimation factor used. For a decimation factor $q$, the subsampled series corresponding to the main analysis can be written as
\begin{equation}
\{(t_{1+kq},x_{1+kq})\}, \qquad k=0,1,2,\ldots,
\end{equation}
considering only indices contained within the original series. The values $q=2$, 4, and 12 were used to represent the nominal 60, 30, and 10 FPS conditions, respectively.

In all cases, the original time and position values corresponding to the selected points were left unchanged. No interpolation or smoothing was performed, and no additional observations were generated. Because the original time values were preserved, the labels 120, 60, 30, and 10 FPS represent the nominal sampling rates used to identify the different experimental conditions.

The same main decimation criterion was used in the free-fall and simple-pendulum experiments. Thus, for each realisation, series derived from the same physical event were compared, preventing differences between FPS conditions from being attributed to variations in initial conditions, spatial calibration, trajectory, or tracking procedure.

This design specifically isolates the effect of reducing the number of available temporal observations. However, when the decimation factor is greater than one, different subsampling phases are possible, determined by the point in the original series at which the selection begins. This dependence was subsequently evaluated through a specific analysis for the pendulum at 10 FPS, described in Section~2.7.

\subsection{Free-fall experiment}
The free-fall experiment consisted of ten independent realisations. In each realisation, a ball traveled a vertical distance of 2.0 m between the two reference marks of the setup.

For each realisation, Tracker was used to obtain a series of pairs
\begin{equation}
(t_i,x_i), \qquad i=1,\ldots,N,
\end{equation}
where $t_i$ is time and $x_i$ is the vertical position of the object. The vertical coordinate was defined as positive in the direction of fall.

The trajectory was fitted using the quadratic model
\begin{equation}
x(t)=A+Bt+Ct^2,
\end{equation}
where $A$ represents a position constant, $B$ accounts for a possible residual initial velocity, and $C$ contains the contribution associated with the acceleration of motion.

Under the adopted convention for the vertical axis, the experimental measurement of the acceleration due to gravity was obtained as
\begin{equation}
g=2C.
\end{equation}
Using $A$ and $B$ as free parameters avoids artificially imposing a specific initial position or an exactly zero initial velocity.

For each release, the fit was first performed on the original 120 FPS record and subsequently on the series reduced to 60, 30, and 10 FPS using the decimation procedure described above. Thus, measurements corresponding to different FPS conditions within the same realisation originate from the same physical event, so the observed differences are associated with the reduction in temporal sampling density rather than with variations among independent releases.

\subsection{Pendulum experiment}
The second experiment used a simple pendulum of nominal length
\begin{equation}
L=0.50\,\mathrm{m}.
\end{equation}
The main analysis was performed on eight independent realisations. In each realisation, five complete oscillations of the pendulum were recorded, and Tracker was used to obtain the time evolution of the horizontal position of the mass.

The oscillation period was determined by a global fit of the time series to a damped oscillatory function,
\begin{equation}
x(t)=x_0+A e^{-\beta t}\cos(\omega t+\phi),
\end{equation}
where $x_0$ represents the equilibrium position, $A$ the initial amplitude, $\beta$ the damping coefficient, $\omega$ the angular frequency, and $\phi$ the initial phase.

From the fitted parameter $\omega$, the period was calculated as
\begin{equation}
T=\frac{2\pi}{\omega}.
\end{equation}
The acceleration due to gravity was estimated using the simple-pendulum relation in the small-oscillation approximation,
\begin{equation}
g=\frac{4\pi^2L}{T^2}.
\end{equation}
The initial angle of the oscillations was approximately $10^\circ$, so the small-angle expression was used consistently as the main model for all realisations and temporal sampling conditions.

For each realisation, the fitting procedure was applied to the original 120 FPS record and subsequently to the series reduced to 60, 30, and 10 FPS by the temporal decimation described above. Thus, measurements corresponding to the different FPS conditions originate from the same physical realisation and differ only in the temporal sampling density of the data used in the fit.

\subsection{Evaluation of precision and accuracy}
For each experiment and temporal sampling condition, a set of individual measurements of the acceleration due to gravity, $g_i$, was obtained. From these measurements, the mean, standard deviation across realisations, relative error with respect to the reference value, and root mean square error (RMSE) were calculated.

The mean estimate of $g$ was calculated as
\begin{equation}
\bar g=\frac{1}{N}\sum_{i=1}^{N}g_i,
\end{equation}
where $N$ is the number of realisations considered in each experiment.

The dispersion across realisations was characterised by the sample standard deviation,
\begin{equation}
s_g=\sqrt{\frac{1}{N-1}\sum_{i=1}^{N}(g_i-\bar g)^2},
\end{equation}
which was used as an indicator of precision. According to the metrological terminology adopted in this work, lower dispersion among repeated measurements corresponds to greater precision (Joint Committee for Guides in Metrology, 2012).

To evaluate the closeness of the measurements to the expected value, the reference value
\begin{equation}
g_{\mathrm{ref}}=9.81\,\mathrm{m\,s^{-2}}
\end{equation}
was adopted. This value is used as a common reference for comparing the different experimental conditions and is not intended to represent a local determination of the acceleration due to gravity.

The deviation of the mean estimate from the reference value was quantified using the percentage relative error,
\begin{equation}
\varepsilon_g=\frac{|\bar g-g_{\mathrm{ref}}|}{g_{\mathrm{ref}}}\times100,
\end{equation}
so that lower values of $\varepsilon_g$ indicate closer agreement of the mean estimate with the reference value.

As an overall measure of the error of the individual measurements, the root mean square error was also calculated,
\begin{equation}
\mathrm{RMSE}=\sqrt{\frac{1}{N}\sum_{i=1}^{N}(g_i-g_{\mathrm{ref}})^2}.
\end{equation}
The RMSE simultaneously incorporates the deviation of the mean from the reference value and the dispersion across realisations. Using the sample definition of $s_g$, it can be expressed as
\begin{equation}
\mathrm{RMSE}^2=(\bar g-g_{\mathrm{ref}})^2+\frac{N-1}{N}s_g^2.
\end{equation}
For this reason, the RMSE is interpreted as an overall measure of measurement performance with respect to $g_{\mathrm{ref}}$, rather than as an exclusive measure of precision or accuracy.

The same metrics, $\bar g$, $s_g$, $\varepsilon_g$, and RMSE, were calculated for the free-fall and simple-pendulum experiments under the four temporal sampling conditions considered: 120, 60, 30, and 10 FPS. These metrics distinguish between a mean estimate close to the reference value and a set of individually consistent measurements.

\subsection{Comparison criterion}
The free-fall and simple-pendulum experiments were compared using the same four temporal sampling conditions: 120, 60, 30, and 10 FPS. For each condition, the same statistical metrics were considered in order to maintain a common evaluation criterion for both systems.

Precision was compared using the standard deviation $s_g$, where lower values indicate lower dispersion across realisations and therefore greater precision. The closeness of the mean estimate to the reference value was evaluated using the percentage relative error $\varepsilon_g$, so that lower values correspond to closer agreement between $\bar g$ and $g_{\mathrm{ref}}$. As a complementary measure, the RMSE was used to quantify the overall error of the individual measurements with respect to the reference value, incorporating both the displacement of the mean and the dispersion across realisations.

Thus, the comparison between both experiments was based on the quantities
\begin{equation}
\bar g,\qquad s_g,\qquad \varepsilon_g,\qquad \mathrm{RMSE},
\end{equation}
calculated equivalently for each FPS condition.

This criterion allows a direct comparison of the effect of reducing the temporal sampling rate on precision, closeness to the reference value, and overall performance of the measurements of $g$, while avoiding reliance on internal fit uncertainties that may have different physical and statistical meanings in each experiment.

\subsection{Sensitivity analysis of the subsampling phase}
The main analysis described above uses a single decimation phase, determined by the first point of each record. However, for a decimation factor $q$, there are $q$ possible sequences containing approximately the same number of observations but beginning at different points in the original series.

To determine whether the stability observed in the pendulum at low temporal sampling rate could depend on a particularly favourable choice of this phase, an additional analysis was performed for the 10 FPS condition. Because this condition is obtained from the original 120 FPS record using a decimation factor $q=12$, all 12 possible subsampling phases were evaluated.

For each realisation, the series
\begin{equation}
\{(t_{r+1+12k},x_{r+1+12k})\},\qquad r=0,1,\ldots,11,
\end{equation}
were constructed, where $r$ identifies the initial offset of the sequence and $k=0,1,2,\ldots$ runs over the available points within each record.

Exactly the same fitting procedure defined for the pendulum was applied to each of these series. Thus, for each of the eight realisations, 12 estimates of $g$ were obtained, associated exclusively with different subsampling phases of the same original trajectory.

Sensitivity to phase was characterised using the standard deviation of the 12 estimates obtained for each realisation and the total range
\begin{equation}
\Delta g_{\mathrm{phase}}=\max_r(g_r)-\min_r(g_r).
\end{equation}
In addition, the mean of the estimates corresponding to the 12 phases was calculated for each realisation. These quantities were used to determine whether the choice of decimation phase appreciably modifies the overall estimate of $g$ at 10 FPS and to compare the magnitude of this effect with the dispersion among experimental realisations.

This sensitivity analysis was used solely as a robustness test of the pendulum results at 10 FPS. The main comparison among 120, 60, 30, and 10 FPS remained based on the same decimation phase defined from the first point of each record for both experiments.

\section{Results}
\subsection{Consolidated results}
Table~\ref{tab:results} summarises the results obtained for the free-fall and simple-pendulum experiments under the four temporal sampling conditions considered. Ten independent realisations were analysed for free fall, whereas the eight realisations included in the main analysis were used for the simple pendulum. For each condition, the table presents the mean estimate of the acceleration due to gravity, $\bar g$, the standard deviation across realisations, $s_g$, the relative error of the mean with respect to $g_{\mathrm{ref}}=9.81\,\mathrm{m\,s^{-2}}$, $\varepsilon_g$, and the root mean square error, RMSE.

\begin{table}[ht]
\centering
\caption{Consolidated results for the measurement of $g$ in the free-fall and simple-pendulum experiments. The standard deviation $s_g$ is used as an indicator of precision, $\varepsilon_g$ quantifies the closeness of $\bar g$ to the reference value, and the RMSE represents an overall measure of the error of the individual measurements with respect to $g_{\mathrm{ref}}$.}
\label{tab:results}
\begin{tabular}{lrrrrr}
\toprule
Experiment & FPS & $\bar g$ (m/s$^2$) & $s_g$ (m/s$^2$) & $\varepsilon_g$ (\%) & RMSE (m/s$^2$)\\
\midrule
Free fall & 120 & 9.8854 & 0.2212 & 0.7684 & 0.2230\\
& 60 & 9.8525 & 0.2753 & 0.4334 & 0.2646\\
& 30 & 9.7858 & 0.3779 & 0.2463 & 0.3593\\
& 10 & 9.7879 & 0.3399 & 0.2257 & 0.3232\\
Simple pendulum & 120 & 9.9257 & 0.0348 & 1.1789 & 0.1201\\
& 60 & 9.9251 & 0.0348 & 1.1734 & 0.1196\\
& 30 & 9.9248 & 0.0349 & 1.1703 & 0.1194\\
& 10 & 9.9247 & 0.0368 & 1.1688 & 0.1197\\
\bottomrule
\end{tabular}
\end{table}

The results show different behaviours for the two systems. In free fall, reducing FPS produces appreciable changes both in the dispersion across realisations and in the position of the mean estimate relative to the reference value. For the simple pendulum, by contrast, all four metrics remain essentially constant over the entire range from 10 to 120 FPS.

\subsection{Mean measurement of \texorpdfstring{$g$}{g} and dispersion}
\begin{figure}[ht]
\centering
\includegraphics[width=0.72\linewidth]{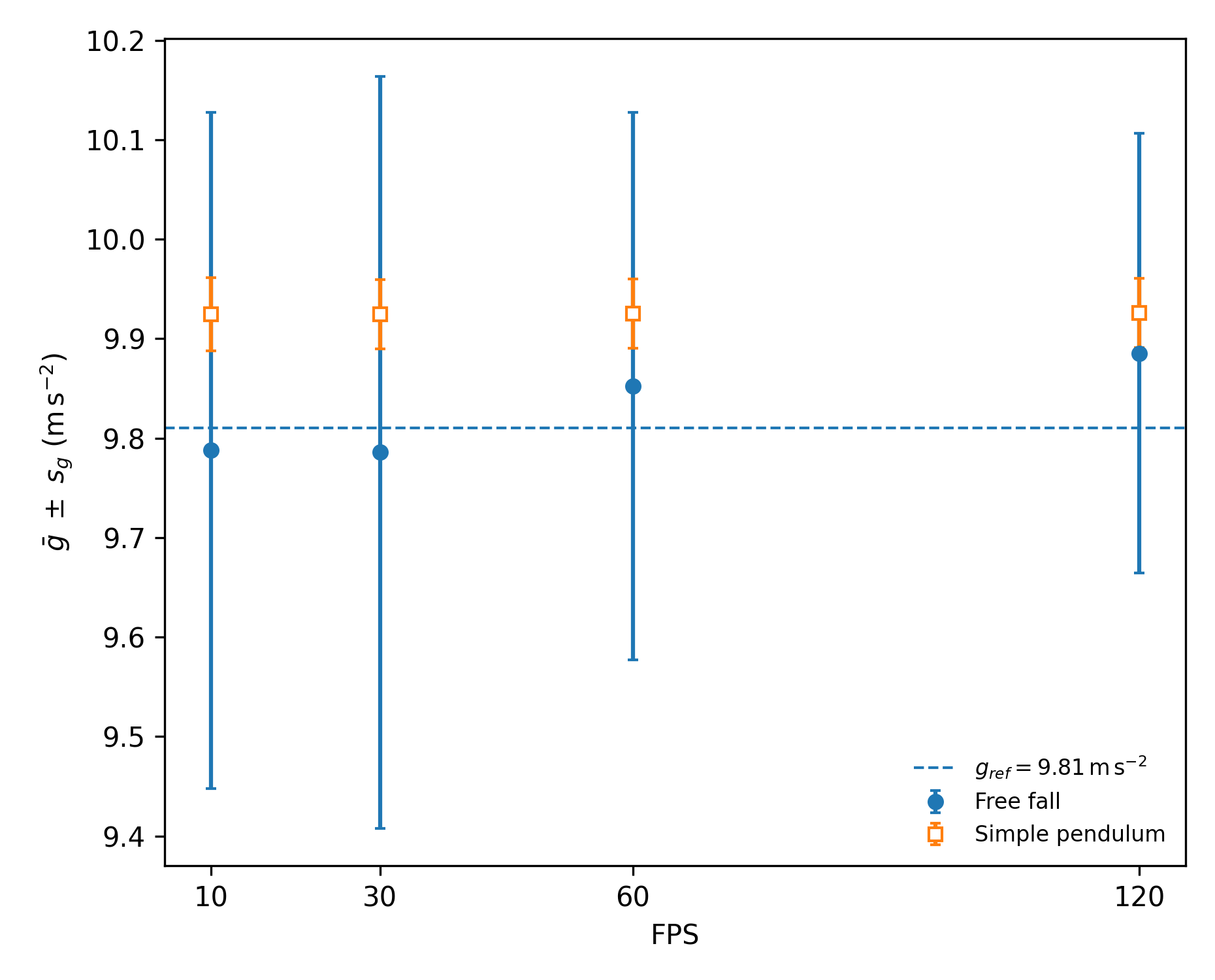}
\caption{Mean acceleration due to gravity for the free-fall and simple-pendulum experiments at 10, 30, 60, and 120 FPS. The vertical bars represent the standard deviation $s_g$ across independent realisations. The horizontal dashed line corresponds to $g_{\mathrm{ref}}=9.81\,\mathrm{m\,s^{-2}}$.}
\label{fig:gmean}
\end{figure}

Figure~\ref{fig:gmean} simultaneously presents the mean estimate of $g$ and the dispersion across realisations for both experiments. The symbols represent $\bar g$, and the vertical bars correspond to one standard deviation, $s_g$. The horizontal dashed line indicates the reference value $g_{\mathrm{ref}}=9.81\,\mathrm{m\,s^{-2}}$.

In free fall, the lowest dispersion was obtained at 120 FPS, with $s_g=0.2212\,\mathrm{m\,s^{-2}}$. As the temporal sampling rate was reduced, the dispersion increased to $0.2753\,\mathrm{m\,s^{-2}}$ at 60 FPS, reached $0.3779\,\mathrm{m\,s^{-2}}$ at 30 FPS, and decreased slightly to $0.3399\,\mathrm{m\,s^{-2}}$ at 10 FPS. The overall behaviour therefore indicates greater dispersion as FPS is reduced, although the variation is not strictly monotonic.

The evolution of $\bar g$ does not follow the same behaviour. At 120 FPS, $\bar g=9.8854\,\mathrm{m\,s^{-2}}$ was obtained, whereas at 60 FPS the value decreased to $9.8525\,\mathrm{m\,s^{-2}}$. At 30 and 10 FPS, the values were $9.7858$ and $9.7879\,\mathrm{m\,s^{-2}}$, respectively. Consequently, the 10 and 30 FPS conditions have the means closest to $g_{\mathrm{ref}}$, despite exhibiting greater dispersion than the 120 FPS condition.

For the simple pendulum, the dependence on FPS is considerably smaller. The mean varies only between $9.9247$ and $9.9257\,\mathrm{m\,s^{-2}}$, while $s_g$ remains between $0.0348$ and $0.0368\,\mathrm{m\,s^{-2}}$. The differences among the four conditions are therefore small compared with those observed in free fall.

\subsection{Relationship between precision, closeness to the reference value, and overall error}
\begin{figure}[ht]
\centering
\includegraphics[width=0.72\linewidth]{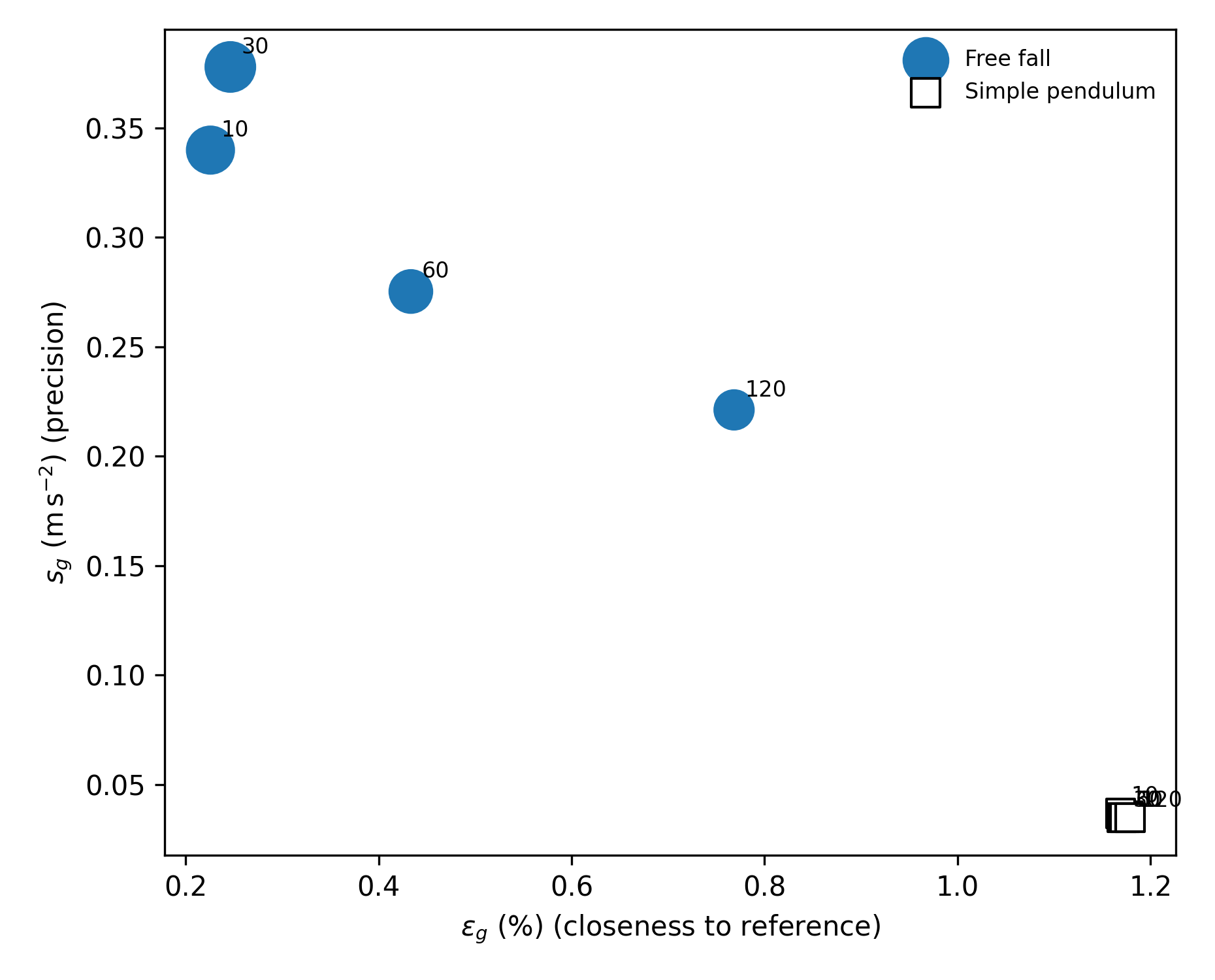}
\caption{Relationship between precision, closeness to the reference value, and overall error for the measurement of $g$. The horizontal axis shows the relative error $\varepsilon_g$, the vertical axis the standard deviation $s_g$, and marker size represents the RMSE. The numbers identify the temporal sampling rate in FPS. Lower values on both axes and smaller markers correspond to better simultaneous performance according to the metrics considered.}
\label{fig:map}
\end{figure}

Figure~\ref{fig:map} combines the three metrics used to characterise measurement performance. The horizontal axis represents the relative error $\varepsilon_g$, which quantifies the closeness of the mean estimate to the reference value, whereas the vertical axis represents $s_g$, used as an indicator of precision. The size of each marker is proportional to the RMSE. The numbers associated with each point indicate the corresponding temporal sampling rate in FPS.

In free fall, a clear separation is observed between precision and closeness to the reference value. The 120 FPS condition has the lowest dispersion, $s_g=0.2212\,\mathrm{m\,s^{-2}}$, and the lowest RMSE, $0.2230\,\mathrm{m\,s^{-2}}$, but at the same time it has the largest relative error of the mean, $\varepsilon_g=0.7684\%$. In contrast, the 10 and 30 FPS conditions show the smallest relative errors, $0.2257\%$ and $0.2463\%$, respectively, but have larger dispersion and RMSE values.

The simple pendulum occupies a compact region of the graph. The values of $s_g$ remain close to $0.035\,\mathrm{m\,s^{-2}}$, the relative errors remain around $1.17\%$, and the RMSE remains around $0.12\,\mathrm{m\,s^{-2}}$. Reducing the frame rate from 120 to 10 FPS therefore produces only very small variations in the three metrics.

The comparison between the two systems shows that a higher temporal sampling rate does not necessarily improve all measurement characteristics simultaneously. In free fall, increasing FPS is associated with lower dispersion and lower RMSE, but not with closer agreement of the mean with the reference value. For the simple pendulum, on the other hand, none of the metrics shows an appreciable dependence on FPS within the range studied.

\subsection{Sensitivity to the subsampling phase in the pendulum}
To determine whether the stability observed at 10 FPS could depend on the phase selected during temporal decimation, all 12 possible subsampling phases from each original 120 FPS record were analysed according to the procedure described in Section~2.7.

The variability introduced exclusively by the choice of phase was small compared with the dispersion across realisations. At 10 FPS, the mean standard deviation of the estimates of $g$ across the 12 phases was
\begin{equation}
s_{\mathrm{phase}}=0.00555\,\mathrm{m\,s^{-2}},
\end{equation}
whereas the standard deviation across the eight realisations in the original 120 FPS condition was
\begin{equation}
s_g=0.03478\,\mathrm{m\,s^{-2}}.
\end{equation}
As a reference for scale, the mean standard deviation associated with phase is approximately one sixth of the standard deviation observed across realisations at 120 FPS. These two quantities characterise different sources of variability and should not be interpreted as equivalent estimates of the same uncertainty.

Averaging, for each realisation, the estimates obtained from the 12 possible phases at 10 FPS yielded
\begin{equation}
\bar g_{10,\mathrm{phases}}=9.925654\,\mathrm{m\,s^{-2}},
\end{equation}
with a standard deviation across realisations of
\begin{equation}
s_g=0.034784\,\mathrm{m\,s^{-2}}.
\end{equation}
These values are effectively indistinguishable from those obtained at 120 FPS,
\begin{equation}
\bar g_{120}=9.925656\,\mathrm{m\,s^{-2}}.
\end{equation}
This agreement should be interpreted descriptively, since the 12 subsampling phases are obtained from the same original 120 FPS record and therefore do not constitute statistically independent datasets. Nevertheless, the small dispersion observed among phases shows that the estimate obtained at 10 FPS does not critically depend on a particular choice of decimation phase.

The total range of the variation due to phase,
\begin{equation}
\Delta g_{\mathrm{phase}}=\max_r(g_r)-\min_r(g_r),
\end{equation}
varied between approximately $0.0027$ and $0.0415\,\mathrm{m\,s^{-2}}$, depending on the realisation, with a mean value
\begin{equation}
\overline{\Delta g}_{\mathrm{phase}}=0.0183\,\mathrm{m\,s^{-2}}.
\end{equation}
Although this effect is not zero, its magnitude remains bounded and does not alter the overall conclusion obtained from comparing the different frame rates.

Furthermore, in all eight realisations, among the 12 possible 10 FPS phases there was at least one phase whose estimate of $g$ was closer to the reference value than the estimate obtained at 120 FPS and, simultaneously, at least one other phase whose estimate was farther away. Defining
\begin{equation}
\delta=|g-g_{\mathrm{ref}}|,
\end{equation}
then, for each realisation,
\begin{equation}
\delta_{10,\min}<\delta_{120}<\delta_{10,\max}.
\end{equation}
This result shows that a higher frame rate reduces the dependence of the estimate on the particular choice of temporal data points, but does not by itself guarantee that the resulting value will be closer to the reference value.

\subsection{Comparative summary}
The two experiments respond differently to a reduction in temporal sampling rate. In free fall, decreasing FPS is generally associated with an increase in dispersion across realisations and in the RMSE. However, the mean estimate does not follow the same trend: the 10 and 30 FPS conditions provide values of $g$ closer to $g_{\mathrm{ref}}$ than the 120 FPS condition.

For the simple pendulum, by contrast, $\bar g$, $s_g$, $\varepsilon_g$, and RMSE remain essentially unchanged between 10 and 120 FPS. The additional analysis of the 12 possible subsampling phases at 10 FPS further shows that this stability is not an accidental result associated with the choice of a particular decimation phase. The variability associated with phase was small compared with the dispersion observed across realisations and did not produce an appreciable systematic shift in the mean estimate of $g$.

Taken together, these results show that the effect of frame rate depends on the dynamical system and on the procedure used to estimate the physical parameter. A higher temporal sampling rate may reduce dispersion and increase measurement robustness to sampling without necessarily guaranteeing closer agreement of the mean estimate with the reference value. The physical interpretation of these behaviours is developed in Section~4.

\section{Discussion}
The results show that reducing the temporal sampling rate does not affect the two procedures used to estimate $g$ in the same way. In free fall, decreasing FPS produces appreciable changes in the dispersion across realisations and in the RMSE. In the simple pendulum, by contrast, $\bar g$, $s_g$, $\varepsilon_g$, and RMSE remain essentially constant over the range from 10 to 120 FPS. Therefore, the influence of temporal sampling rate depends on the dynamical system and on the procedure used to extract the acceleration due to gravity.

The free-fall experiment shows particularly clearly that precision and closeness to the reference value do not necessarily evolve together. The 120 FPS condition has the lowest dispersion, with $s_g=0.2212\,\mathrm{m\,s^{-2}}$, and the lowest RMSE, $0.2230\,\mathrm{m\,s^{-2}}$. However, its relative error, $\varepsilon_g=0.7684\%$, is larger than those obtained at 60, 30, and 10 FPS. The 10 and 30 FPS conditions have means closer to $g_{\mathrm{ref}}$, but simultaneously show considerably greater dispersion. Consequently, judging experimental performance solely from the proximity of $\bar g$ to the reference value would lead, in this case, to an incomplete interpretation.

The behaviour of the RMSE is consistent with this observation. Because this metric incorporates both the displacement from the reference value and the dispersion of the individual measurements, the 120 FPS condition has the lowest overall error even though its mean is not the closest to $g_{\mathrm{ref}}$. The precision-versus-closeness map shown in Figure~\ref{fig:map} makes this separation visible: the lower-FPS free-fall conditions move toward smaller values of $\varepsilon_g$, but simultaneously toward larger values of $s_g$ and RMSE.

This behaviour can be related to the way $g$ is obtained in the free-fall experiment. The measurement comes from the quadratic coefficient of a trajectory recorded over a relatively short, non-repeating interval. Reducing FPS directly decreases the number of observations available to determine the curvature of $x(t)$. Under these conditions, small variations in ball localisation, spatial calibration, or tracking carry greater relative weight in the fit. The results are consistent with this scenario: dispersion generally increases as temporal sampling density is reduced, although the behaviour between 30 and 10 FPS is not strictly monotonic.

The pendulum presents a different situation. Under the four conditions studied, $s_g$ remains approximately between $0.0348$ and $0.0368\,\mathrm{m\,s^{-2}}$, while $\varepsilon_g$ remains around $1.17\%$. The measurements are therefore considerably more tightly clustered than in free fall, but remain displaced from $g_{\mathrm{ref}}$. This result provides another example showing that high precision does not necessarily imply closer agreement with the reference value.

The systematic offset observed in the pendulum estimates should not be attributed directly to frame rate. Because it remains essentially constant for all FPS conditions, it is compatible with systematic contributions common to the experimental procedure, such as determination of the effective pendulum length, spatial calibration, the small-angle approximation, or tracking of the mass position. The aim of the present study is not to identify these contributions separately, but to determine how the estimate changes when the temporal sampling density of the same record is modified.

The stability observed in the pendulum can be associated with the periodic nature of the motion and with the procedure used to determine its period. Unlike free fall, where each instant corresponds to a unique stage of a short trajectory, the pendulum record contains several cycles of the same motion. The information needed to determine the period is therefore repeated throughout the record. Even after a substantial reduction in temporal sampling rate, the series retain sufficient information about the overall periodicity so that the resulting estimates of $T$, and consequently of $g$, change very little.

In periodic motion, reducing FPS also introduces a potential dependence on the subsampling phase. Decimation changes not only the number of observations per cycle but also the relative positions within the oscillation at which the selected points occur. In principle, different phases of the same decimation could produce different temporal distributions even when the total number of observations were nearly the same.

The additional analysis performed at 10 FPS makes it possible to evaluate this possibility directly. Considering all 12 possible subsampling phases, the mean standard deviation associated exclusively with phase was $s_{\mathrm{phase}}=0.00555\,\mathrm{m\,s^{-2}}$. As a reference for scale, this value is approximately one sixth of the standard deviation observed across realisations at 120 FPS, although the two quantities represent different sources of variability and should not be interpreted as equivalent measures of the same uncertainty.

Averaging over the 12 phases yielded $\bar g_{10,\mathrm{phases}}=9.925654\,\mathrm{m\,s^{-2}}$, a value effectively identical to $\bar g_{120}=9.925656\,\mathrm{m\,s^{-2}}$. This agreement should be interpreted descriptively because the subsampling phases originate from the same original 120 FPS record and do not constitute statistically independent datasets. Nevertheless, the small variability observed among phases indicates that the stability obtained at 10 FPS does not critically depend on a particularly favourable selection of decimation phase.

The range of variation due to phase is not strictly zero and depends on the realisation. However, its magnitude remains bounded and does not produce an appreciable systematic shift in the overall estimate of $g$. This result is consistent with the fact that the angular frequency is obtained through a global fit over several complete oscillations. Even at 10 FPS, the selected points remain distributed over multiple phases of the motion, so the period can be estimated from information repeated over several cycles.

Moreover, because the oscillation period does not correspond to an integer number of 0.1 s sampling intervals, the selected points do not systematically recur at the same positions in each cycle. Consequently, reducing the temporal sampling rate decreases the number of available observations and moderately increases sensitivity to the sampling phase, but does not generate an appreciable systematic bias in the estimate of $g$ under the conditions studied.

The phase analysis also reinforces the distinction between precision and closeness to the reference value. For all eight realisations considered, among the 12 possible phases at 10 FPS there was at least one phase with an estimate of $g$ closer to $g_{\mathrm{ref}}$ than that obtained at 120 FPS and, simultaneously, at least one phase with an estimate farther away. This shows that higher temporal resolution can make the result more robust to point selection and reduce its variability without necessarily guaranteeing a smaller deviation from the reference value.

The two experiments demonstrate that the number of frames per second is not, by itself, a measure of the quality of a video-analysis experiment. Its relevance depends on the temporal scale of the phenomenon, the total duration of the record, the presence or absence of repeated temporal information, and the method used to estimate the physical parameter. A short transient trajectory, such as the free fall studied here, is more sensitive to a reduction in the number of observations than a periodic motion recorded over several cycles.

Consequently, the results do not support the idea that increasing FPS necessarily improves both precision and agreement with the reference value. In free fall, a higher temporal sampling rate is clearly associated with lower dispersion and lower RMSE, but not with a reduction in the relative error of the mean. In the pendulum, increasing FPS from 10 to 120 produces no appreciable changes in the global metrics considered, although a higher temporal sampling rate reduces the dependence of the estimate on the subsampling phase.

From a methodological standpoint, these results also show the importance of jointly evaluating $\bar g$, $s_g$, $\varepsilon_g$, and RMSE. The mean provides the central value obtained, $s_g$ characterises precision across realisations, $\varepsilon_g$ quantifies the closeness of that mean to the reference value, and the RMSE provides an overall measure incorporating both contributions. Considering these quantities simultaneously avoids automatically identifying a mean close to the expected value with a higher-quality measurement and allows the effect of temporal sampling rate on video analysis to be characterised more completely.

\section{Conclusions}
This work evaluated the effect of frame rate on the experimental measurement of the acceleration due to gravity using two different dynamical systems: free fall and a simple pendulum. The comparison was performed using the same nominal temporal sampling conditions, 120, 60, 30, and 10 FPS, and the same statistical metrics for both experiments.

The results show that the effect of FPS depends on the system analysed and on the procedure used to estimate $g$. In free fall, reducing the temporal sampling rate generally produced an increase in dispersion across realisations and in the RMSE. The 120 FPS condition had the lowest standard deviation and the lowest RMSE, whereas the 10 and 30 FPS conditions yielded mean values of $g$ closer to the reference value. Therefore, greater precision and lower overall error did not necessarily imply closer agreement of the mean with $g_{\mathrm{ref}}$.

A different behaviour was observed for the simple pendulum. The mean estimate of $g$, the standard deviation across realisations, the relative error, and the RMSE remained essentially constant between 10 and 120 FPS. The additional analysis of the 12 possible subsampling phases at 10 FPS further showed that this stability does not depend on a particularly favourable choice of decimation phase. The variability associated with phase was small compared with the dispersion observed across realisations, and its consideration did not produce an appreciable systematic shift in the mean estimate of $g$.

The comparison between the two experiments therefore shows that increasing FPS does not guarantee improvement in all measurement characteristics simultaneously. In free fall, a higher temporal sampling rate is mainly associated with reduced dispersion across realisations and lower overall error, but not necessarily with closer agreement with the reference value. In the pendulum, increasing FPS produces no appreciable changes in the global metrics considered within the range studied, although it reduces the sensitivity of the estimate to the subsampling phase.

The difference observed between the two systems can be associated with the temporal nature of the motions and with the method used to obtain $g$. Free fall is a short-duration transient phenomenon in which reducing FPS directly decreases the amount of information available to determine the curvature of the trajectory. The pendulum, by contrast, is a periodic system whose temporal information is repeated over several cycles, allowing a stable measurement of the period even when the number of observations is substantially reduced.

The results also indicate that FPS should not be used in isolation as a quality criterion in video-analysis experiments. The duration of the record, the transient or periodic nature of the motion, the temporal distribution of the observations, and the procedure used to estimate the physical parameter must be considered jointly. In particular, a higher temporal sampling density may increase precision or robustness to sampling without necessarily implying an estimate closer to the reference value.

Consequently, the appropriate temporal sampling rate should be selected according to the characteristics of the phenomenon and the physical quantity to be determined. Under the conditions studied, a larger number of frames provides additional information and may improve certain measurement characteristics, but does not by itself guarantee greater precision and accuracy simultaneously.

\end{document}